\documentclass[12pt]{article}

\usepackage{amsbsy}
\usepackage{wasysym}
\usepackage{color}
\usepackage{graphicx}
\usepackage{amssymb}

\begin{document}

\title{Universal structure of blow-up \\ in 1D conservation laws}

\author{Alexei A. Mailybaev\thanks{Instituto Nacional de Matem\'atica Pura e Aplicada -- IMPA, Rio de Janeiro, Brazil. 
Institute of Mechanics, Moscow State University, Russia. E-mail: a.mailybaev@gmail.com, alexei@impa.br.}}
\date{\today}

\maketitle

\begin{abstract}
We discuss universality properties of blow-up of a classical (smooth) solutions of conservation laws in one space dimension. It is shown that the renormalized wave profile tends to a universal function, which is independent both of initial conditions and of the form of a conservation law. This property is explained in terms of the renormalization group theory. A solitary wave appears in logarithmic coordinates of the Fourier space as a counterpart of this universality. Universality is demonstrated in two examples: Burgers equation and dynamics of ideal polytropic gas. 
\end{abstract}

\noindent\textbf{Keywords:} blow-up; conservation law; universality; renormalization group; solitary wave; inviscid Burgers equation

\section{Introduction}

Conservation laws constitute the foundation of modern physics. They are linear when describing single or several elementary particle systems, and become nonlinear for macroscopic systems, motivating the vast research area of nonlinear wave dynamics. The important class of conservation laws corresponds to idealized models, when the flux functions depend on state variables only (neglecting derivative terms corresponding to dissipation etc.), see, e.g., \cite{dafermos2010hyperbolic,whitham1974linear}. In these models, smooth initial conditions typically give rise to a blow-up (singularity) in finite time followed by formation of a shock wave. 
Development of such finite-time singularities is a classical subject described in every book on nonlinear waves, e.g., \cite{alinhac1995blowup,dafermos2010hyperbolic,john1990nonlinear,whitham1974linear}.

It is well understood that scalar conservation laws in one space dimension capture a lot of what happens in general systems. In this respect, the inviscid Burgers (or Hopf) equation 
\begin{equation}
\frac{\partial u}{\partial t}
+u\frac{\partial u}{\partial x} = 0
\label{H.1}
\end{equation}
plays a special role. Solution of this equation is easily constructed by the method of characteristics. Each characteristic carries a constant value of $u$, and the blow-up occurs when characteristics cross. 

In this paper we show that the wave profile $u(t,x)$ creates a universal ``core" just before the shock formation, as a consequence of a cusp catastrophe in $(t,x,u)$ space. This core is described, up to scaling symmetry, by an analytic function, which is independent of initial conditions as well as of the flux function. We explain the universality using the renormalization group approach, and discuss its relation to more sophisticated universality phenomena described by the theory of renormalization group like critical phenomena of second-order phase transitions \cite{wilson1974renormalization} among many others \cite{feigenbaum1978quantitative,kadanoff2011}. Finally, we show that in logarithmic coordinated of the Fourier transformed function $u(t,k)$, the blow-up is mapped to a stable solitary wave moving with constant speed. As an example, we describe universal structure of shock formation in ideal polytropic gas.

\section{Universal structure of blow-up}

First, let us consider the inviscid Burgers equation (\ref{H.1})
with smooth initial condition $u(t_0,x) = u_0(x)$. Its solution $u(t,x)$ constructed by the method of characteristics has the implicit form
\begin{equation}
x = x_0+u_0(x_0)(t-t_0),\quad
u = u_0(x_0),
\label{H.2}
\end{equation}
where $x_0$ is an auxiliary variable. 
For spatial derivative of $u(t,x)$, we have
\begin{equation}
\frac{\partial u}{\partial x}
= \frac{\partial u/\partial x_0}{\partial x/\partial x_0} 
= \frac{u_0'(x_0)}{1+u_0'(x_0)(t-t_0)}.
\label{H.3}
\end{equation}
The denominator vanishes at $t = t_0-1/u_0'(x_0)$.
This yields the well-known result that the classical solution blows-up along the characteristic with the minimum negative value of $u_0'(x_0)$, followed by the formation of a shock wave in a weak solution. 

We choose the origin of time and space  so that the blow-up singularity appears at $t = x = 0$, and consider the classical solution in the interval $t_0 \le t < 0$. Also, we can take $u = 0$ at the singularity, which can be achieved by the transformation $x \mapsto x-u_0(0)(t-t_0)$ and $u \mapsto u+u_0(0)$, which leaves (\ref{H.1}) invariant. In this case, the initial condition $u_0(x)$ satisfies 
\begin{equation}
u_0(0) = 0,\quad t_0 = 1/u_0'(0) < 0,\quad u_0''(0) = 0,\quad u_0'''(0) > 0,
\label{H.4}
\end{equation}
which are the blow-up conditions at $t = x = u = 0$. 
Using (\ref{H.4}) in (\ref{H.2}), for small $x_0$, we obtain
\begin{equation}
x = ut-\frac{u_0'''(0)}{6u'(0)}x_0^3+o(x_0^3),\quad
u = u'_0(0)x_0+o(x_0).
\label{H.5}
\end{equation}
Equivalently,
\begin{equation}
x = ut-cu^3+o(u^3),\quad
c = \frac{u_0'''(0)}{6(u'(0))^4} > 0.
\label{H.6}
\end{equation}
In particular, at $t = 0$, we obtain the well-known singular dependence $u \sim -x^{1/3}$. 

Consider now the renormalization function
\begin{equation}
u_\lambda(t,x) = \mathcal{G}_{\lambda}u(t,x) 
\equiv \lambda^{1/3}u(\lambda^{-2/3}t,\,\lambda^{-1} x).
\label{H.7}
\end{equation}
It is easy to see that $u_\lambda$ is a new (scaled) solution of (\ref{H.1}), so (\ref{H.7}) represents one of the symmetries of (\ref{H.1}). Multiplying both sides of (\ref{H.6}) by $\lambda$ and making the substitution $t \mapsto \lambda^{-2/3}t$, $x \mapsto \lambda^{-1}x$ yields the equation for $u_\lambda(t,x)$ as
\begin{equation}
x = u_\lambda t-cu_\lambda^3+\lambda\, o\!\left((\lambda^{-1/3}u_\lambda)^3\right).
\label{H.8}
\end{equation}
The last (correction) term contains powers $(\lambda^{-1/3}u_\lambda)^n$ with $n > 3$. Hence, it vanishes in the limit of large $\lambda$,
and (\ref{H.8}) takes the exact form 
\begin{equation}
x = w t-cw^3
\label{H.10}
\end{equation}
for the limiting function
\begin{equation}
w(t,x) = \lim_{\lambda \to \infty}u_\lambda(t,x).
\label{H.9}
\end{equation}

\begin{figure*}
\centering \includegraphics[width=0.49\textwidth]{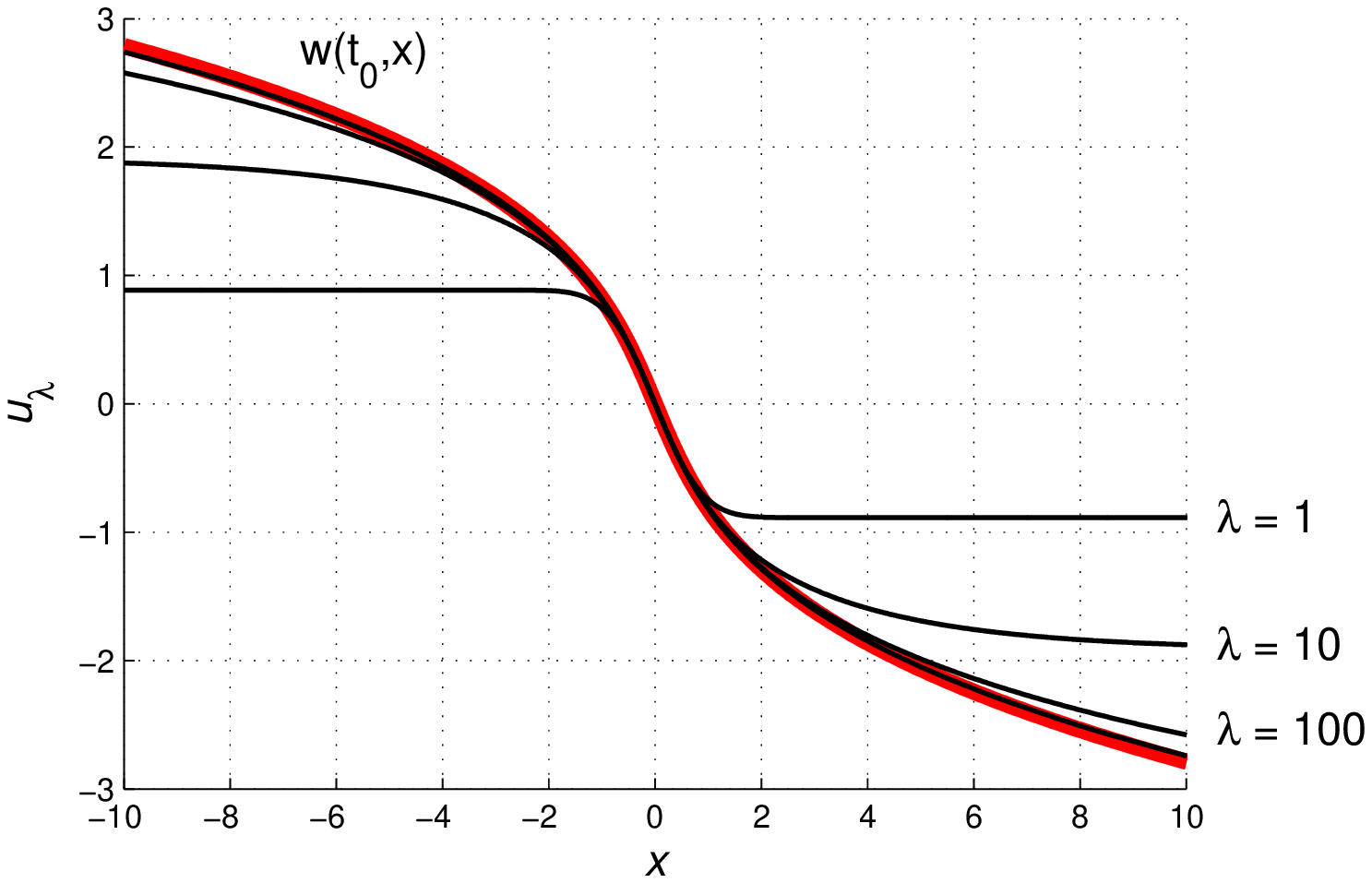}
\centering \includegraphics[width=0.49\textwidth]{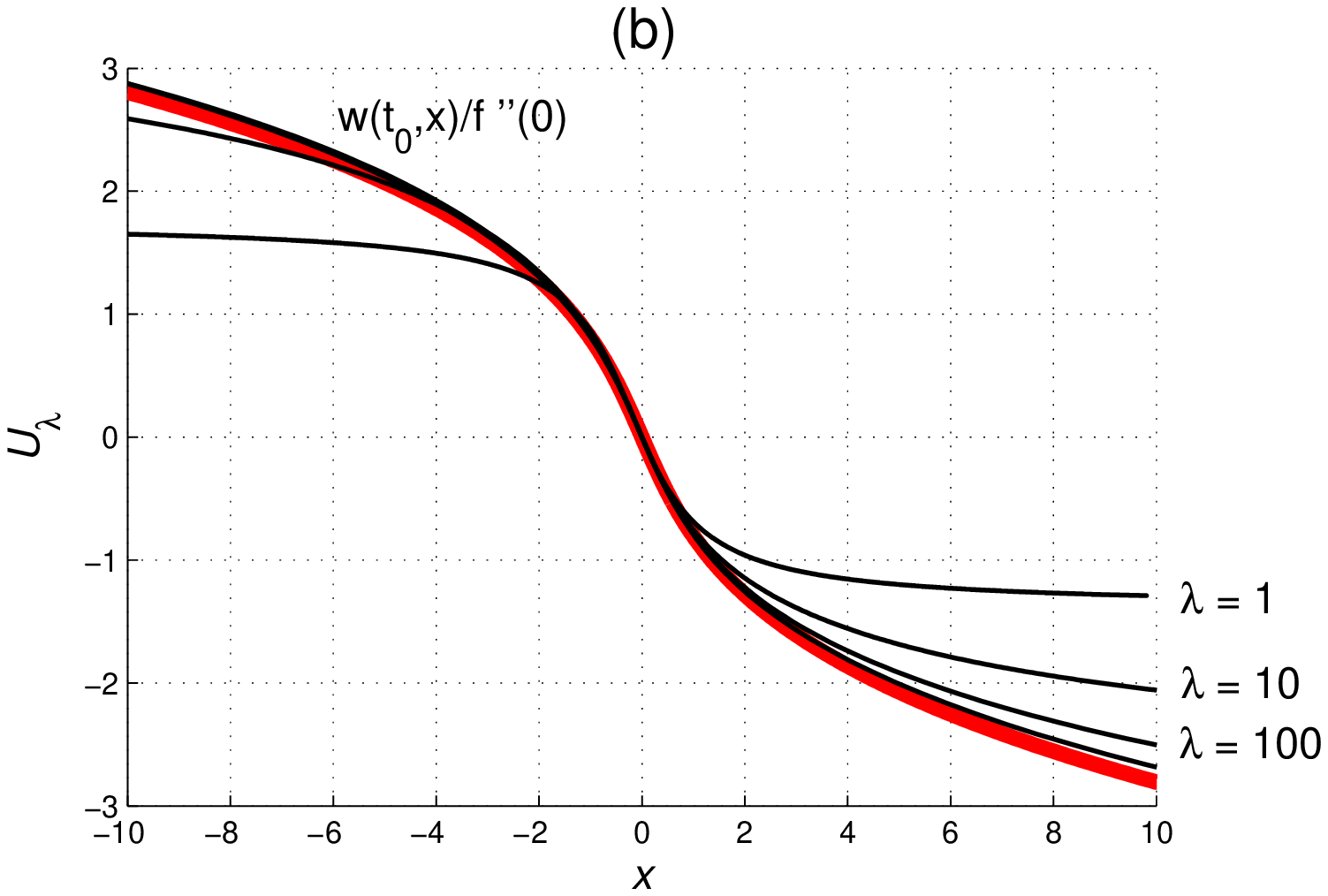}
\caption{(a) Convergence of scaled profiles $u_\lambda(t_0,x)$ (thin black curves) to the universal function $w(t_0,x)$ (bold red curve)  for the inviscid Burgers equation (\ref{H.1}). The initial condition is $u_0(x) = -(\sqrt{\pi}/2)\textrm{erf}\,x$ at $t_0 = -1$, and $\lambda = 1,\,10,\,10^2,\,10^3$. (b) Similar convergence results (\ref{C.3}) for blow-up of a simple wave in ideal polytropic gas. The bold red curve represents the scaled universal function $w(t_0,x)/f''(0)$, and thin back curves are the scaled profiles $U_\lambda(t_0,x) = \mathcal{G}_\lambda U(t_0,x)$ corresponding to the density variation $U$ determined by (\ref{G.1}), (\ref{G.2c}), (\ref{G.2e}).} 
\label{fig1}
\end{figure*}

Equation (\ref{H.10}) determines the universal function $w(x,t)$ for $t \le 0$, which is independent of initial conditions. Note that different values of the coefficient $c$ correspond to $w(t,x)$ determined up to scaling $\sqrt{c}w(t,x/\sqrt{c})$, which is a symmetry of the Burgers equation (\ref{H.1}). The function $w(t,x)$ is a self-similar solution, 
\begin{equation}
\mathcal{G}_\lambda w = w.
\label{R.1}
\end{equation}
A numerical example of convergence of $u(t,x)$ to the universal function $w(t,x)$ is shown in Fig.~\ref{fig1}a. 

The same function $w(t,x)$ describes the limiting universal behavior for a general scalar conservation law
\begin{equation}
\frac{\partial U}{\partial t}
+\frac{\partial f(U)}{\partial x} = 0
\label{C.1}
\end{equation}
with an analytic flux function $f(U)$ and initial condition $U_0(x)$ at $t = t_0$. This fact can be shown as follows. 
Solution of equation (\ref{C.1}) has the form 
\begin{equation}
x = x_0+f'(U_0(x_0))(t-t_0),\quad
U = U_0(x_0),
\label{C.2b}
\end{equation}
which reduces to solution (\ref{H.2})
of the Burgers equation by the substitution $u = f'(U)$. We assume that the coordinates for time, space and state are chosen such that (\ref{H.4}) holds for $u_0(x) = f'(U_0(x))$, and $u = f'(U)$ is locally invertible as $U = g(u)$ with $g(0) = f'(0) = 0$. This implies the blow-up at $t = x = 0$ with $U = u = 0$. We will use the relation
\begin{equation}
\mathcal{G}_{\lambda}u^n(t,x)
= \lambda^{1/3}u^n(\lambda^{-2/3}t,\,\lambda^{-1} x) 
= \lambda^{(1-n)/3}u_{\lambda}^n(t,x).
\label{C.2}
\end{equation}
Expanding $g(u)$ in Taylor series and using (\ref{H.9}), (\ref{C.2}), we see that all terms with $n > 1$ vanish for large $\lambda$. Since $g(0) = 0$ and $g'(0) = 1/f''(0)$, the remaining terms yield
\begin{equation}
\lim_{\lambda\to\infty}\mathcal{G}_{\lambda}U(t,x) 
= \lim_{\lambda\to\infty}\mathcal{G}_{\lambda}g(u(t,x)) 
= w(t,x)/f''(0).
\label{C.3}
\end{equation}

Conditions (\ref{H.9}) and (\ref{C.3}) with the function $w(t,x)$ given by (\ref{H.10}) demonstrate strong universal character of the blow-up of a classical solution for a scalar 1D conservation law. We see that, when the singular point is approached, a part of the wave profile $u(t,x)$ takes an absolutely universal form independent both of the initial condition and of the flux function (up to the scaling transformation). In particular, derivatives of all orders turn out to be universal for the scaled solution near the singular point. 

Note that expressions (\ref{H.6})--(\ref{H.9}) relate the blow-up with a cusp catastrophe in the space $(t,x,u)$, see \cite{arnoldcatastrophe}, so that the universality of the blow-up represents a general property of this catastrophe.
Since the cusp catastrophe is a generic phenomenon, the universality results are extended directly to systems of coupled conservation laws in one space dimension, for which a general solution similar to (\ref{C.2b}) does not exist.  

As an example, let us consider formation of a shock wave in a simple wave solution for one-dimensional flow of ideal polytropic gas. The density $\rho(t,x)$ in this wave is described implicitly by
\begin{equation}
\frac{x-x_0}{t-t_0} = \frac{\gamma+1}{\gamma-1}\sqrt{A\gamma}\rho_0^{(\gamma-1)/2}(x_0),\quad 
\rho = \rho_0(x_0),
\label{G.1}
\end{equation}
where $\rho_0(x)$ is the initial condition at $t = t_0$, 
see, e.g., \cite{courant1977supersonic}. We will use the values $\gamma = 5/3$, $A = 3/5$, and $\rho_0(x) = 2-\arctan x$. 
Then expressions (\ref{G.1}) take the form
\begin{equation}
\begin{array}{l}
x = x_0+4(2-\arctan x_0)^{1/3}(t-t_0),\\[5pt]
\rho = 2-\arctan x_0,
\end{array}
\label{G.1b}
\end{equation}
which can be written in the form (\ref{C.2b}) for the density variation in moving frame
\begin{equation}
U(t,x) = \rho(t,x+x_1+v_1(t-t_0))-\rho_1
\label{G.2c}
\end{equation}
with
\begin{equation}
\begin{array}{c}
f'(U) = 4(U+\rho_1)^{1/3}-v_1,\\[3pt]
U_0(x) = 2-\arctan (x+x_1)-\rho_1.
\end{array}
\label{G.2}
\end{equation}
The quantities $t_0$, $x_1$, $\rho_1$ and $v_1$ 
are chosen such that $f'(0) = 0$ and the function 
\begin{equation}
u_0(x) = f'(U_0(x)) = 4(2-\arctan(x+x_1))^{1/3}-v_1
\label{G.2d}
\end{equation}
satisfies conditions (\ref{H.4}) corresponding to blow-up at $t = x = 0$ with $U = 0$. Values of these quantities (with first three decimal digits) are
\begin{equation}
t_0 = -1.155,\ 
x_1 = 0.183,\ 
\rho_1 = 1.818,\ 
v_1 = 4.882.
\label{G.2e}
\end{equation} 
Convergence (\ref{C.3}) of the scaled wave profile $U_\lambda(t,x) = \mathcal{G}_\lambda U(t,x)$ to the universal function is demonstrated in Fig.~\ref{fig1}b. 

\section{Renormalization group approach}

The universality just described can also be explained using the renormalization group approach. For this purpose, we consider $\mathcal{G}_\lambda$ in (\ref{H.7}) as an operator acting in the space of solutions of the Burgers equation (\ref{H.1}) with initial conditions satisfying (\ref{H.4}). Then the system dynamics can be seen as an action of $\mathcal{G}_\lambda$ combined with the scaling of time, space and state. This operator defines a differentiable group with the property $\mathcal{G}_{\lambda_1}\mathcal{G}_{\lambda_2} = \mathcal{G}_{\lambda_1+\lambda_2}$. The universal function $w(t,x)$ represents a stationary point (\ref{R.1}) of the renormalization group operator. Our analysis showed that this stationary point (more precisely, a set of stationary points $w(t,x)$ determined by (\ref{H.10}) up to a scaling constant $c$) is asymptotically stable in the sense of Lyapunov (for $\lambda$ considered as ``time"). From this property, the universality result (\ref{H.9}) follows. 

In the case of a general conservation law (\ref{C.1}), the function $U_\lambda(t,x) = \mathcal{G}_\lambda U(t,x)$ is a solution for a conservation law with a different flux function  $f_\lambda(U) = \lambda^{2/3}f(\lambda^{-1/3}U)$. Thus, $\mathcal{G}_\lambda$ is a renormalization group operator acting in the functional space $(U,f) \mapsto (U_\lambda,f_\lambda)$ of solutions and fluxes. The universality (\ref{C.3}) of the blow-up is explained by the fact that $\mathcal{G}_\lambda$ has the asymptotically stable stationary point $(U,f) = (w(t,x),U^2/2)$ corresponding to the universal solution (\ref{H.10}) of the Burgers equation (\ref{H.1}). 

The role of the renormalization group operator $\mathcal{G}_\lambda$ is similar to those in other, much more sophisticated theories. For the inner scale, $x \sim ut \sim t^{3/2}$ in (\ref{H.6}), the wave dynamics is governed by the universal function, which is a stationary point of $\mathcal{G}_\lambda$. This is analogous, e.g., to the stationary point of the renormalization group operator, which determines critical phenomena in second-order phase transitions~\cite{wilson1974renormalization}. At larger spatial scales, there is no universality and the solution depends on the initial condition $u_0(x)$ as well as on the flux function $f(u)$. This is analogous, in turn, to the phenomenological Landau theory of second-order phase transitions valid at larger (though still small) deviations of temperature from a critical value~\cite{landau1980statistical}.
 
\section{Blow-up as a solitary wave in wavenumber space}

Finite time blow-up implies rapid increase of solution in a short wavelength range. We will see now that the universality of the blow-up in $x$-space induces a solitary wave moving with constant speed to large wave numbers in logarithmic coordinates. Consider the Fourier transform in $x$-space of the solution, $u(t,k) = \mathcal{F}_x[u(t,x)]$.
Since $u(t,x)$ is real, we have $u(t,-k) = u^*(t,k)$, and we  assume $k \ge 0$ in the analysis that follows. For Fourier transformed functions, the relation (\ref{H.9}) yields 
\begin{equation}
w(t,k) = \lim_{\lambda \to \infty}u_\lambda(t,k),
\label{N.1}
\end{equation}
with the $k$-space renormalization group operator 
\begin{equation}
u_\lambda(t,k) = \mathcal{G}_\lambda u(t,k) \equiv 
\lambda^{4/3}u(\lambda^{-2/3}t,\,\lambda k).
\label{N.2}
\end{equation}
Expression (\ref{N.2}) can be checked using Fourier transform of (\ref{H.7}).
Note that the function $w(t,x)$ grows as $w \sim x^{1/3}$ for large $x$, see (\ref{H.10}), and its Fourier transform (regularized by adding a small imaginary part to $x$) behaves as $w \sim k^{-4/3}$ for small $k$. 

Let us consider the wave profile $u(t,k)$ transformed to logarithmic coordinates 
\begin{equation}
\tau = -\log (t/t_0),\quad
\xi = \log k,
\label{N.4}
\end{equation}
which we denote by the same letter $u(\tau,\xi)$. Here the blow-up at $t = 0$ corresponds to $\tau \to \infty$. 
Using (\ref{N.2}), we find
\begin{equation}
u_\lambda(\tau,\xi) = e^{4a/3}
u\left(\tau+\frac 23a,\,\xi+a\right),\quad
a = \log\lambda.
\label{N.5a}
\end{equation}
Because of (\ref{N.1}), we have 
\begin{equation}
e^{4a/3}u\left(\frac 23a,\,\xi+a\right) \to w(0,\xi)
\quad \textrm{as} \quad a \to \infty,
\label{N.5b}
\end{equation}
where we put $\tau = 0$, and $w(\tau,\xi)$ denotes the universal function $w(t,k)$ written in coordinates (\ref{N.4}). Now we substitute $a$ by $3\tau/2$ and write (\ref{N.5b}) in the form 
\begin{equation}
u\left(\tau,\,\xi+\frac 32\tau\right) \to e^{-2\tau}w(0,\xi)
\quad \textrm{as} \quad \tau \to \infty.
\label{N.5bb}
\end{equation}
Therefore, $u(\tau,\xi)$ forms a wave in the $\xi$-space for large $\tau$ with the profile described by the universal function $w(0,\xi)$. This wave moves with the constant speed $3/2$ and decays as $e^{-2\tau}$. 

\begin{figure}
\centering \includegraphics[width=0.6\textwidth]{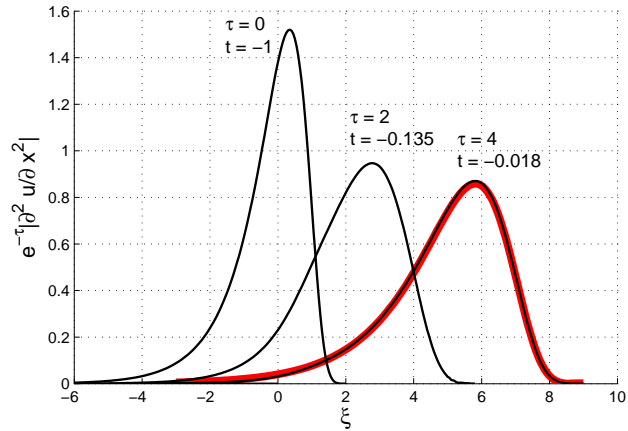}
\caption{Profiles of the scaled second derivative $e^{-\tau}|\partial^2u/\partial x^2|$ in logarithmic Fourier coordinates $(\tau,\xi)$ defined by (\ref{N.4}) for the solution shown in Fig.~\ref{fig1}(a). Initial condition corresponds to $\tau = 0$, and the blow-up corresponds to $\tau \to \infty$. The profile forms a solitary wave traveling with constant speed $3/2$, which reflects the universality of the blow-up in $x$- and $k$-spaces. The bold red curve shows the exact limiting shape of the wave determined by the universal function $w(t,x)$.}
\label{fig2}
\end{figure}

Fourier transforms of derivatives $u^{(n)}(t,x) = \partial^n u/\partial x^n$ have the form $u^{(n)} (t,k) = (ik)^nu(t,k)$. Using logarithmic coordinates $(\tau,\xi)$, derivations similar to (\ref{N.4})--(\ref{N.5bb}) yield
\begin{equation}
u^{(n)}\left(\tau,\,\xi+\frac 32\tau\right) \to i^ne^{(3n/2-2)\tau}e^{n\xi}w(0,\xi)
\quad \textrm{as} \quad \tau \to \infty.
\label{N.5bbb}
\end{equation}
Recall that $w \sim k^{-4/3} = e^{-4\xi/3}$ for $\xi \to -\infty$ (small $k$). Thus, $e^{n\xi}w(0,\xi) \to 0$ as $\xi \to -\infty$ when $n > 4/3$. In this case the wave derivative $u^{(n)}$ forms a solitary wave in the space $(\tau,\xi)$, which moves with speed $3/2$, grows exponentially as $e^{(3n/2-2)\tau}$, and has the universal shape determined by $e^{n\xi}w(0,\xi)$. Formation of such a wave is shown in Fig.~\ref{fig2} for the second derivative, $n = 2$. Note that a similar wave describes a finite-time blow-up in shell models of turbulence, see \cite{dombre1998intermittency}.  

\section{Conclusion}

We showed that solutions of conservation laws in one space dimension have universal structure when approaching the finite-time singularity (blow-up). The limiting wave profile is described by a scaled universal function in $x$-space, or by a solitary wave moving with constant speed in logarithmic coordinates of (wave number) $k$-space. The universal function is independent both of initial conditions and of specific expressions for fluxes of conserved quantities.  
The universal properties are confirmed numerically in Figs.~\ref{fig1} and \ref{fig2}. Experimental observation does not seem to be difficult due to large variety of nonlinear waves described by conservation laws in physical system. 
The obtained results can be extended to higher dimensional spaces, in particular 2D and 3D, where the universality can be used as an effective tool for identification of blow-up singularities. This extension will be published elsewhere.

\section*{Acknowledgment} 
The author is grateful to C.M.~Dafermos, E.A.~Kuznetsov and D.~Marchesin for useful discussions related to this paper.

\bibliographystyle{plain}
\bibliography{refs}

\end{document}